# Multi-wavelength analysis of the galactic PeVatron candidate LHAASO J2108+5157

**G. Pirola[a,*], J. Juryšek[b], M. Balbo[c], D. Eckert[c], A. Tramacere[c], and R. Walter[c] on behalf of the CTA-LST project**

[a] *Max-Planck-Institut für Physik, Föhringer Ring 6, 80805 München, Germany*
[b] *FZU - Institute of Physics of the Czech Academy of Sciences, Na Slovance 1999/2 182 00, Prague 8, Czechia*
[c] *Department of Astronomy, University of Geneva, Chemin d'Ecogia 16, CH-1290 Versoix, Switzerland*

*E-mail:* lst-contact@cta-observatory.org, gpirolampp@mpg.de

LHAASO J2108+5157 is a recently discovered source, detected in the Ultra-High-Energy band by the LHAASO collaboration. Two molecular clouds were identified in the direction coincident with LHAASO J2108+5157 and, from the spectra reported by LHAASO, there is no sign of an energy cutoff up to 200 TeV. This source makes a promising galactic PeVatron candidate.
In 2021, the Large-Sized Telescope prototype (LST-1) of the Cherenkov Telescope Array (CTA) Observatory, collected about 50 hours of quality-selected data on LHAASO J2108+5157. Through these observations, we managed to compute stringent upper limits on the source emission in the multi-TeV band. Together with the analysis of *XMM-Newton* data and 12 years of *Fermi-LAT data*, we performed a multi-wavelength study of the source, investigating different possible scenarios of particle acceleration.

In this contribution, we will present the results of the analysis, as well as the multi-wavelength modeling, and consequent interpretation of different possible scenarios of emission.

*Speaker

## 1. Introduction

Galactic PeVatrons are one of the still unresolved puzzles of high-energy astrophysics. We expect Cosmic Rays (CRs) in our Galaxy to be accelerated up to PeV energies, but although Supernova Remnants (SNRs) have always been considered the best candidates for Galactic CR acceleration, so far there is no evidence that they can accelerate particles up to few PeV, and the question on how particles are accelerated up to the *knee* of the CR spectrum is still far from being answered (see [1]).

In order to study Galactic hadrons acceleration at PeV energies, it is necessary to look at Very-High-Energy (VHE, between 100 GeV and 100 TeV) and Ultra-High-Energy (UHE, between 100 TeV and 100 PeV) gamma rays: PeV CR protons colliding with the ambient gas produce $\pi^0$ that decay into the gamma-ray photons with about 1/10 of the energy of the progenitor relativistic proton. Studying sources of gamma rays above 100 TeV is then necessary to unveil the mystery of the UHE Galactic CRs. However, the presence of UHE gamma-ray emission does not represent a direct evidence of hadron acceleration: gamma rays can also be produced by electrons and positrons via inverse Compton (IC) scattering on low-energy photons, via bremsstrahlung on atomic nuclei in the surrounding matter, or also by emitting synchrotron radiation when traveling across a magnetic field. In the case of leptonic acceleration, in the VHE range, the dominant contribution comes from IC, which at energies greater than $\sim$ 100 TeV is suppressed due to the Klein-Nishina effect. Nevertheless, IC can still dominate UHE emission in radiation-dominated environments [2, 3].

In 2021, the LHAASO collaboration detected significant gamma-ray emissions at energies up to 1.4 PeV from 12 UHE $\gamma$-ray sources [4]. Among them, LHAASO J2108+5157 is the first one in the UHE band without any known VHE or X-ray counterpart [5]. In the High-Energy (HE) range, between 1 GeV and 500 GeV, 4FGL J2108.0+5155 is a soft point-like source found at an angular distance of 0.13° [5]. Furthermore, a 0.26° upper limit on the source extension with 95% confidence level was found [5]. Although the leptonic hypothesis could not be ruled out, two molecular clouds were identified in the direction coincident with LHAASO J2108+5157, making it a promising candidate for the hadron acceleration scenario [5].

In this contribution, we will present how LST-1 data managed to put stringent upper limits on the source emission in the multi-TeV band. We will also show the results of a multi-instrument analysis including *XMM-Newton* data and 12 years of *Fermi*-LAT data, as well as the multi-wavelength modeling and consequent interpretation of different possible scenarios of emission. The results discussed in this contribution have been recently published [6].

## 2. Observation and data analysis

### 2.1 LST-1 data

Inaugurated on October $10^{th}$ 2018, the prototype for CTA of the Large-Sized Telescope LST-1, located at the *Observatorio del Roque de Los Muchachos* (ORM) in La Palma, is currently going through its commissioning phase.

During the Summer of 2021, LST-1 observed LHAASO J2108+5157 for 91 hours of observations in wobble mode. We selected 49.3 hours of good-quality data, by applying a series of cuts based on trigger rate stability, rate of CR events, and atmospheric transmission. The event reconstruction process was performed using the Python-based pipeline implemented in `lstchain v0.9` [7]. We extracted the signal from the $\theta^2$ distributions using the reflected region method, and although the source was not significantly detected, we found a $3.67\sigma$ point-like excess in the highest energy bin (3-100 TeV). With the `gammapy` package [8], we performed the 1D spectral analysis. We assumed a power law (PL) $\mathrm{d}N/\mathrm{d}E = N_0(E/E_0)^{-\Gamma}$ spectral model: the results of the fit are reported in the table shown in fig 1. We then performed a joint likelihood fit of the LST-1 data and LHAASO flux points, by fitting the data with a power-law with exponential cutoff (ECPL) $\mathrm{d}N/\mathrm{d}E = N_0(E/E_0)^{-\Gamma} exp(-E/E_c)$. Secondly, we performed a maximum-likelihood estimation of the source flux in six energy bins between 100 GeV and 100 TeV, using the ECPL spectral parameters fitted in the previous step. Even if we did not reach a significant source detection, LST-1 data provide strong ULs on the source emission in the TeV band (see fig. 1). Fig. 1 also shows a first attempt of building a significance map using a tool[1] for the creation of an acceptance model from real data, that can be used for radial corrections in `Gammapy` background models (see [6]).

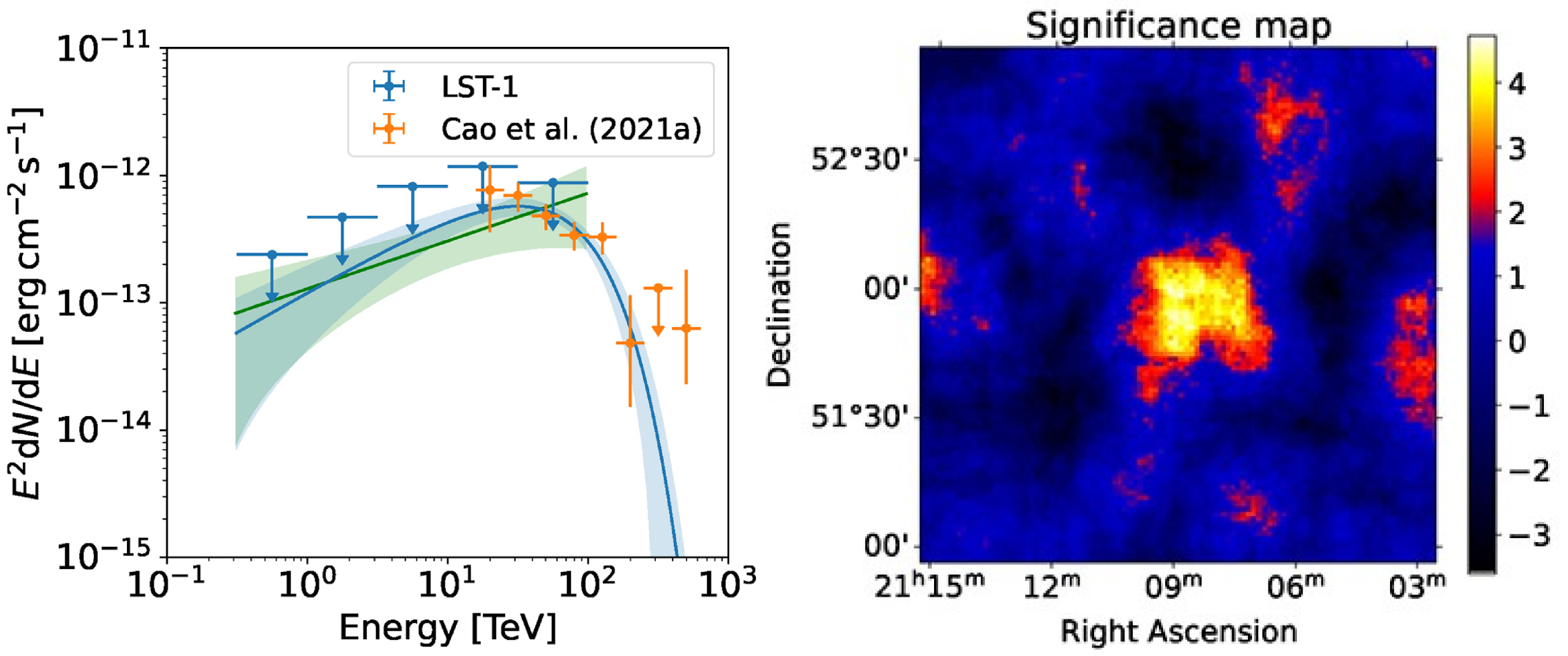

| Data | Spectral model | $N_0$ [$\times 10^{-14}$ cm$^{-2}$ s$^{-1}$ TeV$^{-1}$] | $\Gamma$ | $E_{\rm cutoff}$ [TeV] | $-2\log\mathcal{L}$ |
|---|---|---|---|---|---|
| LST-1 | PL | $8.0 \pm 5.4$ | $1.62 \pm 0.23$ | ... | 5.17 |
| LST-1 + LHAASO | ECPL | $7.6 \pm 4.8$ | $1.37 \pm 0.22$ | $50 \pm 14$ | 7.30 |

**Figure 1:** *Left*: LST-1 measured spectral energy distribution (SED) of LHAASO J2108+5157. The green line and band represent the best-fit PL spectral model of LST-1 data and its statistical uncertainties, whereas the blue line corresponds to the joint likelihood fit of the LST-1 data and LHAASO flux points with an ECPL spectral model. Together with the ECPL model, the plot also shows the 95% confidence level ULs on differential flux [6]. In the table are reported the best-fit parameters for the two aforementioned spectral analyses [6]. *Right*: statistical significance map in a region of $2° \times 2°$ around LHAASO J2108+5157, in the energy range between 3 TeV and 100 TeV. The map was produced using *ring*-background model assuming a fixed index PL spectrum ($\Gamma = 2$) [6].

[1] https://github.com/mdebony/acceptance_modelisation/tree/main/acceptance_modelisation

### 2.2 Fermi-LAT data

We performed a dedicated binned analysis of the region around LHAASO J2108+5157, but differently from the LHAASO Collaboration, which used the 10-year 4FGL-DR2 catalog 4FGL-DR2 catalog, we adopted the more recent 12-year 4FGL-DR3 catalog [9] for the modeling of the sources in the field of view. In addition to the catalog counterpart of LHAASO J2108+5157, 4FGL J2108.0+5155, we also found that the emission above $\approx$ 4 GeV is dominated by a new source, identified with a $4\sigma$ significance and whose spectrum we fitted with a hard PL ($\Gamma = 1.9$). Given its location (coordinates $l = 92.35°$, $b = 2.56°$), this new hard source is not spatially correlated with LHAASO J2108+5157; nevertheless, by including it in the background model, we could better describe the spectrum of 4FGL J2108.0+5155 [6].

### 2.3 *XMM-Newton* data

*XMM-Newton* observed the field surrounding LHAASO J2108+5157, on June 11, 2021, for a total of 13.6 ks [6]. After reducing the data from the European Photon Imaging Camera (EPIC), we extracted images in the soft (0.5-2 keV) and hard (2-7 keV) bands [6]. We found that the eclipsing binary V1061 Cyg (RX J2107.3+5202) is way brighter than all the other X-ray sources in the field of view, but because of its very soft thermal spectrum, it can't be considered as a counterpart of LHAASO J2108+5157. In case the UHE emission is originated by accelerated leptons, we would expect the source to be surrounded by a synchrotron nebula with an angular size of a few arcminutes. By using the public code `pyproffit` [10], we derived the upper limits on the X-ray emission from a putative extended source centered on the coordinates of LHAASO J2108+5157. We assumed a PL spectrum with a photon index of 2 and we computed the ULs for two opposite hypotheses on the (unknown) source distance: nearby and hence completely unabsorbed, or a maximally absorbed emission from a source situated on the other side of the Galaxy. The ULs were computed for different possible extension values of the region, including $6'$ which corresponds with the maximum possible radius precisely derivable from the available data, due to the source position within the relatively small *XMM-Newton* field of view. In order to estimate a flux upper limit for the region corresponding to the 95% extension upper limits of LHAASO J2108+5157 (UHE), which has a radius of $16'$, we scaled the $6'$ upper limits by the ratio of the two radii [6]. Fig. 2 and 3 show the 95% confidence level ULs on the X-ray emission for the absorbed scenario.

### 2.4 Molecular clouds

In [5] they searched the database of molecular clouds in the Galactic plane [11] and found two molecular clouds, whose direction is in proximity of LHAASO J2108+5157: [MML2017]2870 and [MML2017]4607 with distances 1.43 kpc and 3.28 kpc, and masses $3.5 \times 10^4$ $M_\odot$ and $8.5 \times 10^3$ $M_\odot$, respectively. We used $^{12}$CO(1-0) line emission observations (see [12]) to estimate distance and density of the molecular clouds. Consistently with [5], we identified three peaks in the $T_{B(H_2,v)}$ spectrum at $v_1 \approx -11.8\,\mathrm{km\,s^{-1}}$, $v_2 \approx -2.7\,\mathrm{km\,s^{-1}}$ and $v_3 \approx 8.4\,\mathrm{km\,s^{-1}}$. The two negative ones correspond to the centroid velocities of the molecular clouds identified by *M.A. Miville-Deschênes et al.*([11]), but the origin of the last peak remains unknown and it was not considered for the rest of the analysis. From the Gaussian fit over the velocity distributions, we evaluated the column density values for the two molecular clouds [13]. Assuming a spherically symmetrical source emission region, we then estimated the number density of molecular ($H_2$) and neutral hydrogen (HI) for both clouds (see [6]): $n_1 = n_1(\mathrm{HI}) + n_1(\mathrm{H_2}) = 115\,\mathrm{cm}^{-3}$ and $n_2 = 240\,\mathrm{cm}^{-3}$.

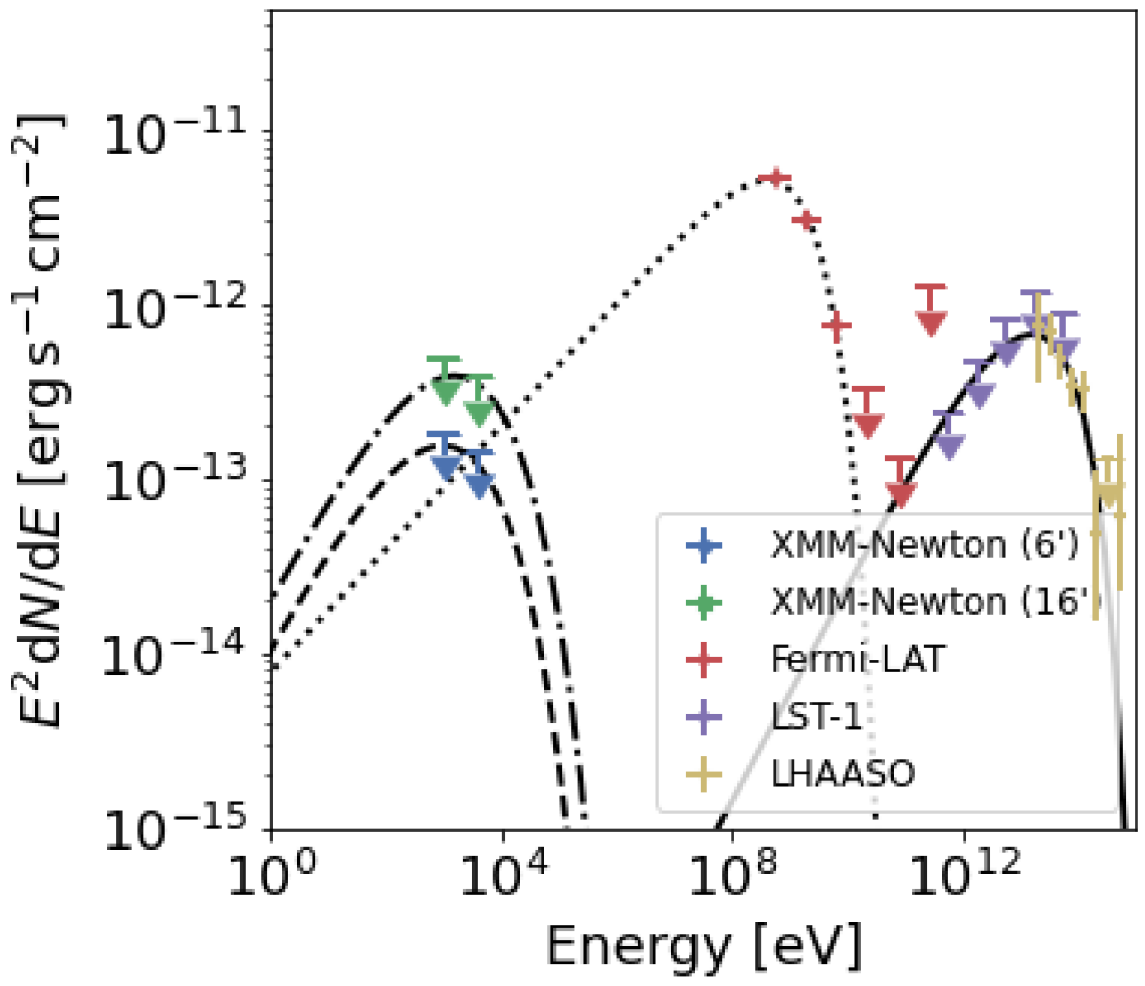

| Parameter | Best fit value | Frozen |
|---|---|---|
| $E_0$ [TeV] | 1 | True |
| $E_{e,min}$ [GeV] | 0.1 | True |
| $E_{e,max}$ [TeV] | 1000 | True |
| $N_0$ [$\times 10^{43}$ TeV$^{-1}$] | $1.7^{+4}_{-1.4}$ | False |
| $E_c$ [TeV] | $100^{+70}_{-30}$ | False |
| $\alpha$ | $1.5 \pm 0.4$ | False |

**Figure 2:** The plot shows the SEDs of LHAASO J2108+5157 together with the leptonic modeling. The solid line represents the best fitting IC-dominated emission of LST-1 and LHAASO data, whereas the dashed and dash-dotted represent the synchrotron radiation of the same electron population for $B = 1.2\,\mu$G and $B = 1.9\,\mu$G, respectively [6]. The dotted line represents a subexponential cutoff PL phenomenological model of a tentative pulsar, fitted on the *Fermi*-LAT data. The table contains the best-fit parameters of the ECPL electron distribution, where $\alpha$ is the spectral index, $E_0$ is the energy scale, $E_c$ is the cutoff energy and $N_0$ the normalization factor [6].

## 3. Discussion of possible emission scenarios

### 3.1 Leptonic emission scenario

We used the `naima`[2] package [14] to test the hypothesis of the leptonic origin of the multi-TeV emission. We assumed an ECPL electron distribution producing gamma rays via IC interaction with Cosmic Microwave Background (CMB) and Far Infrared Radiation (FIR) of the dust. The best-fitting model and parameters are shown in fig. 2. The very constraining ULs on X-ray emission from the synchrotron nebula impose a very strong limit on the source magnetic field ($B \leq 1.2\,\mu$G for a 6′-radius extended region and $B \leq 1.9\,\mu$G for the 16′ case). Such values are not easily reconcilable with the Pulsar Wind Nebula (PWN) scenario, which, as we know, describes most of the currently known VHE sources [15]. Nevertheless, the constraints on the Magnetic field are consistent with the TeV halo class of objects [16], where values even below Galactic background level can be expected [17]. Furthermore, assuming a Geminga-like diffusion coefficient and source distance $d \geq 2$ kpc, the possible extension seen at the ultra-high energies is compatible with the size of a typical TeV halo object. Looking at the GeV range, IC-dominated radiation of a single electron population cannot explain the soft emission of 4FGL J2108.0+5155. We therefore advanced the hypothesis that the GeV emission is the signature of a $\gamma$-ray pulsar: we fitted the *Fermi*-LAT data with a phenomenological spectral model, usually adopted for $\gamma$-ray pulsar emission in the HE range, and we evaluated its spin-down power to be in the range of $\dot{E} \approx 10^{34} - 10^{37}$ erg s$^{-1}$ [6]. In the Geminga-like scenario, we found that for a source extension of about 10 pc, the aforementioned lower limit on the source distance imposes a pulsar spin-down power $\dot{E} \geq 10^{35}$ [6]. Given the cutoff energy of 100 TeV, for $B \leq 1.9\,\mu$G, the corresponding cooling time of relativistic electrons is $t_{\rm cool} \approx 20$ kyr. With a spin down power $\dot{E} \geq 10^{36}$ erg s$^{-1}$, the putative pulsar could power the IC PWN emission for any possible distance of the source in the Galaxy [6].

[2] `https://naima.readthedocs.io/en/latest/`

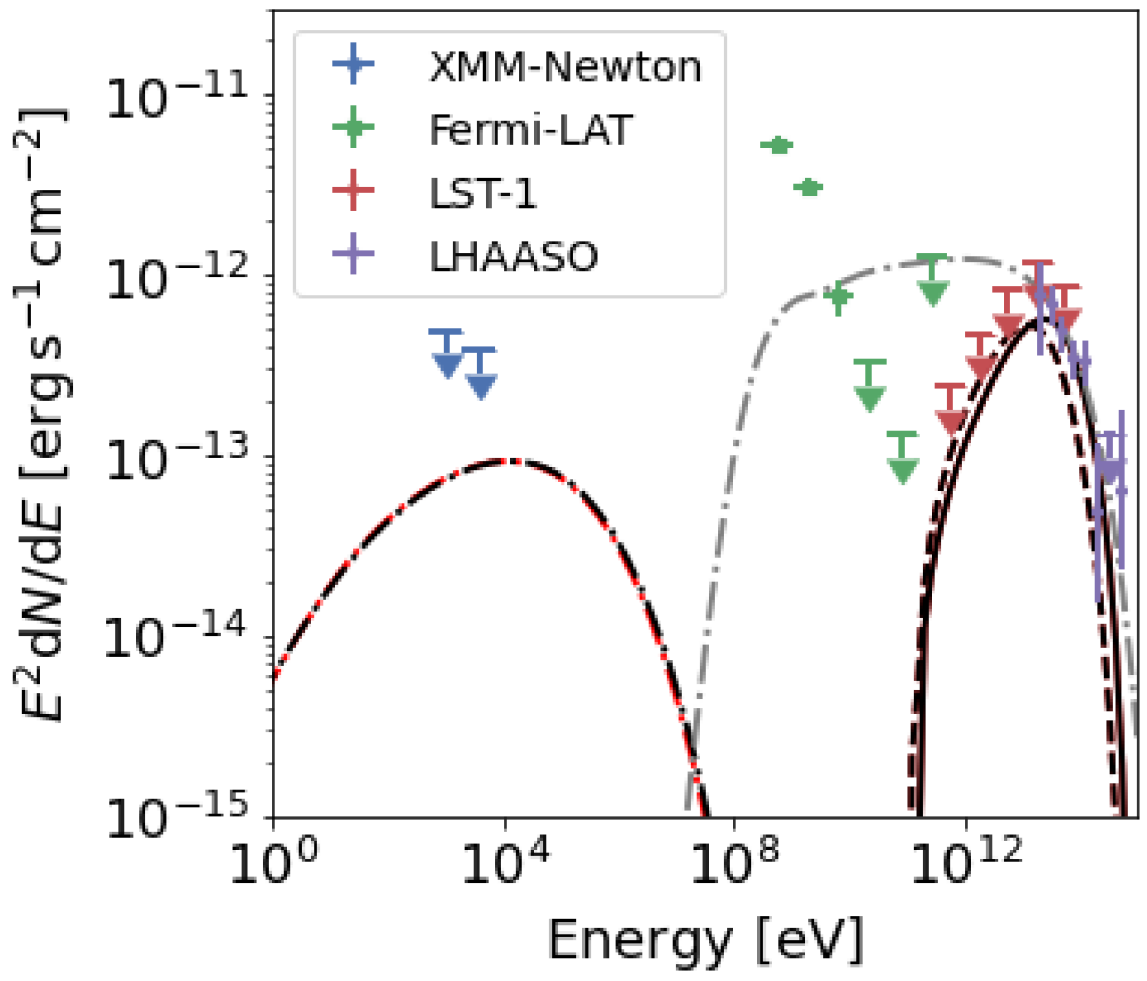

| Parameter | Best fit value | | Frozen |
|---|---|---|---|
| | Cloud 1 | Cloud 2 | |
| $n$ [cm$^{-3}$] | 115 | 240 | True |
| $d$ [kpc] | 3.1 | 2.0 | True |
| R [pc] | 7.1 | 4.5 | True |
| $\gamma_{\rm min}$ [$\times 10^5$] | $1.6 \pm 0.5$ | $1.6 \pm 5$ | False |
| $\gamma_{\rm max}$ [$\times 10^6$] | 1.0 | 1.0 | True |
| B [mG] | $9 \pm 5$ | $\leq 8$ | False |
| $N$ [$\times 10^{-15}$ cm$^{-3}$] | $4 \pm 1$ | $3 \pm 1$ | False |
| $\alpha$ | 2.75 | 2.75 | True |

**Figure 3:** The plot shows the SEDs of LHAASO J2108+5157 together with the hadronic modeling. The solid line represents the best-fit $\pi^0$ decay emission model. The Grey dashed line represents $\pi^0$ decay emission model with proton spectral index $\alpha = 2$ shown for reference [5]. The table contains the best-fit parameters of a $\pi^0$-decay-dominated VHE-UHE scenario of emission for the two molecular clouds. The injected protons are assumed to be distributed according to an ECPL with a $\gamma$-factor in the range ($\gamma_{min}$, $\gamma_{max}$), a $\gamma_{cut}$ cutoff, an $\alpha$ spectral index, and an N total numeric density (S. Abe et al. (2023) [6]).

### 3.2 Hadronic emission scenario

The presence of two dense molecular clouds, together with the absence of an X-ray counterpart, supports the hadronic scenario hypothesis: UHE gamma rays are the decay product of $\pi^0$, resulting from inelastic collisions of hadronic relativistic CR with the hadrons contained in the clouds. We investigated the SNR hypothesis, already proposed by [5]. Assuming a $\pi^0$ decay-dominated origin of the UHE emission, we used the `jetset`[3] package to fit the LST-1 and LHAASO flux points. Fig. 3 shows the $\pi^0$ decay emission model with $\alpha = 2.$ and $\gamma_{\rm min} = 1$ proposed by [5], which we found to be ruled out by the *Fermi*-LAT and LST-1 ULs [6]. Therefore we framed our model into the phenomenological scenario whose base assumption is that of a middle-aged SNR, in which the hard spectrum is generated by protons escaping the shock and illuminating one of the molecular clouds [6]. We tested the hypothesis by modeling the proton distribution with a PL with fixed index ($\alpha = 2.75$), whereas the value of $\gamma_{\rm min}$ is set free in the fit [6]. In this scenario, $\gamma_{\rm min}$ mimics the energy break investigated in [18], and $\alpha$ represents the index of the escaped proton distribution [6]. Fig. 3 shows the best-fit models found for both molecular clouds. Assuming the $\pi^0$ emitted photons are not absorbed by the source, the total neutrino flux resulting from $\pi^{+/-}$ decay is comparable with the gamma-ray flux in the TeV range [6].

[3]https://jetset.readthedocs.io/en/latest/

It is not possible to explain, with a single-component hadronic model, the HE gamma-ray emission of the *Fermi*-LAT counterpart and the UHE one. The $\Gamma = -3.2$ photon index above 1 GeV resulting from our analysis of *Fermi*-LAT data is be too soft compared to the spectra of old SNRs interacting with dense molecular clouds [19] as it was suggested by [5]. However, if considering the hypothesis of significant CR sea contribution to the HE emission, the hadronic origin of the HE emission cannot be excluded [6].

## 4. Conclusions

We analyzed and combined data from LST-1, *XMM-Newton*, Fermi-LAT and LHAASO, performing a multi-wavelength study of the unidentified UHE gamma-ray source LHAASO J2108+5157. Despite being still in its commissioning phase, LST-1 provided data capable of producing constraining ULs on the source emission in the VHE range. We investigated both leptonic and hadronic emission scenarios. The former suggests that the UHE source may belong to the class of TeV halo objects, which are not very common and may provide important insight into CR particle diffusion within the Galaxy, since they appear to have a diffusion coefficient much lower than that of the interstellar medium. The PWN hypothesis is also supported by the soft HE emission, which, if originating from a pulsar, could provide the relativistic electrons with enough energy to power the UHE emission.

Concerning the hadronic scenario, the ULs placed by LST-1 in the VHE range ruled out the model proposed by [5]. We therefore proposed a different model, under the more complex assumption of protons escaping the shock around an old SNR [18]. Although our model reproduces the observed broad-band SED reasonably, a follow-up analysis is needed to get a better constraint on the spectrum of the escaped protons illuminating the molecular cloud [6].

## Acknowledgments

For the acknowledgments see: `https://www.lst1.iac.es/acknowledgements.html`.

## References

[1] S. Gabici, C. Evoli, D. Gaggero, P. Lipari, P. Mertsch, E. Orlando et al., *International Journal of Modern Physics D* **28** (2019) 1930022 [1903.11584].

[2] G. Vannoni, S. Gabici and F. A. Aharonian, *A&A* **497** (2009) 17 [0803.1138].

[3] M. Breuhaus, J. Hahn, C. Romoli, B. Reville, G. Giacinti, R. Tuffs et al., *ApJ* **908** (2021) L49 [2010.13960].

[4] Z. Cao, F. A. Aharonian, Q. An, L. X. Axikegu, Bai, Y. X. Bai, Y. W. Bao et al., *Nature* **594** (2021) 33.

[5] Z. Cao, F. Aharonian, Q. An, Axikegu, L. X. Bai, Y. X. Bai et al., *ApJl* **919** (2021) L22 [2106.09865].

[6] S. Abe, A. Aguasca-Cabot, I. Agudo, N. Alvarez Crespo, L. A. Antonelli, C. Aramo et al., **673** (2023) A75 [2210.00775].

[7] R. Lopez-Coto, T. Vuillaume, A. Moralejo, F. Cassol, M. Nöthe, D. Morcuende et al., Apr., 2022. 10.5281/zenodo.6458862.

[8] C. Deil, R. Zanin, J. Lefaucheur, C. Boisson, B. Khelifi, R. Terrier et al., in *35th International Cosmic Ray Conference (ICRC2017)*, vol. 301 of *International Cosmic Ray Conference*, p. 766, Jan., 2017, 1709.01751.

[9] Fermi-LAT collaboration, :, S. Abdollahi, F. Acero, L. Baldini, J. Ballet et al., *arXiv e-prints* (2022) arXiv:2201.11184 [2201.11184].

[10] D. Eckert, A. Finoguenov, V. Ghirardini, S. Grandis, F. Kaefer, J. Sanders et al., *The Open Journal of Astrophysics* **3** (2020) 12 [2009.03944].

[11] M.-A. Miville-Deschênes, N. Murray and E. J. Lee, **834** (2017) 57 [1610.05918].

[12] T. M. Dame, D. Hartmann and P. Thaddeus, **547** (2001) 792 [astro-ph/0009217].

[13] A. D. Bolatto, M. Wolfire and A. K. Leroy, **51** (2013) 207 [1301.3498].

[14] V. Zabalza, in *34th International Cosmic Ray Conference (ICRC2015)*, vol. 34 of *International Cosmic Ray Conference*, p. 922, July, 2015, 1509.03319.

[15] H. E. S. S. Collaboration, H. Abdalla, A. Abramowski, F. Aharonian, F. Ait Benkhali, E. O. Angüner et al., *A&A* **612** (2018) A1 [1804.02432].

[16] R. López-Coto, E. de Oña Wilhelmi, F. Aharonian, E. Amato and J. Hinton, *Nature Astronomy* **6** (2022) 199 [2202.06899].

[17] R.-Y. Liu, C. Ge, X.-N. Sun and X.-Y. Wang, *The Astrophysical Journal* **875** (2019) 149.

[18] S. Celli, G. Morlino, S. Gabici and F. A. Aharonian, **490** (2019) 4317 [1906.09454].

[19] Q. Yuan, S. Liu and X. Bi, *ApJ* **761** (2012) 133 [1203.0085].

## Full Author List: CTA-LST project

K. Abe[1], S. Abe[2], A. Aguasca-Cabot[3], I. Agudo[4], N. Alvarez Crespo[5], L. A. Antonelli[6], C. Aramo[7], A. Arbet-Engels[8], C. Arcaro[9], M. Artero[10], K. Asano[2], P. Aubert[11], A. Baktash[12], A. Bamba[13], A. Baquero Larriva[5,14], L. Baroncelli[15], U. Barres de Almeida[16], J. A. Barrio[5], I. Batkovic[9], J. Baxter[2], J. Becerra González[17], E. Bernardini[9], M. I. Bernardos[4], J. Bernete Medrano[18], A. Berti[8], P. Bhattacharjee[11], N. Biederbeck[19], C. Bigongiari[6], E. Bissaldi[20], O. Blanch[10], G. Bonnoli[21], P. Bordas[3], A. Bulgarelli[15], I. Burelli[22], L. Burmistrov[23], M. Buscemi[24], M. Cardillo[25], S. Caroff[11], A. Carosi[6], M. S. Carrasco[26], F. Cassol[26], D. Cauz[22], D. Cerasole[27], G. Ceribella[8], Y. Chai[8], K. Cheng[2], A. Chiavassa[28], M. Chikawa[2], L. Chytka[29], A. Cifuentes[18], J. L. Contreras[5], J. Cortina[18], H. Costantini[26], M. Dalchenko[23], F. Dazzi[6], A. De Angelis[9], M. de Bony de Lavergne[11], B. De Lotto[22], M. De Lucia[7], R. de Menezes[28], L. Del Peral[30], G. Deleglise[11], C. Delgado[18], J. Delgado Mengual[31], D. della Volpe[23], M. Dellaiera[11], A. Di Piano[15], F. Di Pierro[28], A. Di Pilato[23], R. Di Tria[27], L. Di Venere[27], C. Díaz[18], R. M. Dominik[19], D. Dominis Prester[32], A. Donini[6], D. Dorner[33], M. Doro[9], L. Eisenberger[33], D. Elsässer[19], G. Emery[26], J. Escudero[4], V. Fallah Ramazani[34], G. Ferrara[24], F. Ferrarotto[35], A. Fiasson[11,36], L. Foffano[25], L. Freixas Coromina[18], S. Fröse[19], S. Fukami[2], Y. Fukazawa[37], E. Garcia[11], R. Garcia López[17], C. Gasbarra[38], D. Gasparrini[38], D. Geyer[19], J. Giesbrecht Paiva[16], N. Giglietto[20], F. Giordano[27], P. Gliwny[39], N. Godinovic[40], R. Grau[10], J. Green[8], D. Green[8], S. Gunji[41], P. Günther[33], J. Hackfeld[34], D. Hadasch[2], A. Hahn[8], K. Hashiyama[2], T. Hassan[18], K. Hayashi[2], L. Heckmann[8], M. Heller[23], J. Herrera Llorente[17], K. Hirotani[2], D. Hoffmann[26], D. Horns[12], J. Houles[26], M. Hrabovsky[29], D. Hrupec[42], D. Hui[2], M. Hütten[2], M. Iarlori[43], R. Imazawa[37], T. Inada[2], Y. Inome[2], K. Ioka[44], M. Iori[35], K. Ishio[39], I. Jimenez Martinez[18], J. Jurysek[45], M. Kagaya[2], V. Karas[46], H. Katagiri[47], J. Kataoka[48], D. Kerszberg[10], Y. Kobayashi[2], K. Kohri[49], A. Kong[2], H. Kubo[2], J. Kushida[1], M. Lainez[5], G. Lamanna[11], A. Lamastra[6], T. Le Flour[11], M. Linhoff[19], F. Longo[50], R. López-Coto[4], A. López-Oramas[17], S. Loporchio[27], A. Lorini[51], J. Lozano Bahilo[30], P. L. Luque-Escamilla[52], P. Majumdar[53,2], M. Makariev[54], D. Mandat[45], M. Manganaro[32], G. Manicò[24], K. Mannheim[33], M. Mariotti[9], P. Marquez[10], G. Marsella[24,55], J. Martí[52], O. Martinez[56], G. Martínez[18], M. Martínez[10], A. Mas-Aguilar[5], G. Maurin[11], D. Mazin[2,8], E. Mestre Guillen[52], S. Micanovic[32], D. Miceli[9], T. Miener[5], J. M. Miranda[56], R. Mirzoyan[8], T. Mizuno[57], M. Molero Gonzalez[17], E. Molina[3], T. Montaruli[23], I. Monteiro[11], A. Moralejo[10], D. Morcuende[4], A. Morselli[38], V. Moya[5], H. Muraishi[58], K. Murase[2], S. Nagataki[59], T. Nakamori[41], A. Neronov[60], L. Nickel[19], M. Nievas Rosillo[17], K. Nishijima[1], K. Noda[2], D. Nosek[61], S. Nozaki[8], M. Ohishi[2],

Y. Ohtani[2], T. Oka[62], A. Okumura[63,64], R. Orito[65], J. Otero-Santos[17], M. Palatiello[22], D. Paneque[8], F. R. Pantaleo[20], R. Paoletti[51], J. M. Paredes[3], M. Pech[45,29], M. Pecimotika[32], M. Peresano[28], F. Pfeiffle[33], E. Pietropaolo[66], G. Pirola[8], C. Plard[11], F. Podobnik[51], V. Poireau[11], M. Polo[18], E. Pons[11], E. Prandini[9], J. Prast[11], G. Principe[50], C. Priyadarshi[10], M. Prouza[45], R. Rando[9], W. Rhode[19], M. Ribó[3], C. Righi[21], V. Rizi[66], G. Rodriguez Fernandez[38], M. D. Rodríguez Frías[30], T. Saito[2], S. Sakurai[2], D. A. Sanchez[11], T. Šarić[40], Y. Sato[67], F. G. Saturni[6], V. Savchenko[60], B. Schleicher[33], F. Schmuckermaier[8], J. L. Schubert[19], F. Schussler[68], T. Schweizer[8], M. Seglar Arroyo[11], T. Siegert[33], R. Silvia[27], J. Sitarek[39], V. Sliusar[69], A. Spolon[9], J. Strišković[42], M. Strzys[2], Y. Suda[37], H. Tajima[63], M. Takahashi[63], H. Takahashi[37], J. Takata[2], R. Takeishi[2], P. H. T. Tam[2], S. J. Tanaka[67], D. Tateishi[70], P. Temnikov[54], Y. Terada[70], K. Terauchi[62], T. Terzic[32], M. Teshima[8,2], M. Tluczykont[12], F. Tokanai[41], D. F. Torres[71], P. Travnicek[45], S. Truzzi[51], A. Tutone[6], M. Vacula[29], P. Vallania[28], J. van Scherpenberg[8], M. Vázquez Acosta[17], I. Viale[9], A. Vigliano[22], C. F. Vigorito[28,72], V. Vitale[38], G. Voutsinas[23], I. Vovk[2], T. Vuillaume[11], R. Walter[69], Z. Wei[71], M. Will[8], T. Yamamoto[73], R. Yamazaki[67], T. Yoshida[47], T. Yoshikoshi[2], N. Zywucka[39], M. Balbo[69], D. Eckert[69], A. Tramacere[69]

[1]Department of Physics, Tokai University. [2]Institute for Cosmic Ray Research, University of Tokyo. [3]Departament de Física Quàntica i Astrofísica, Institut de Ciències del Cosmos, Universitat de Barcelona, IEEC-UB. [4]Instituto de Astrofísica de Andalucía-CSIC. [5]EMFTEL department and IPARCOS, Universidad Complutense de Madrid. [6]INAF - Osservatorio Astronomico di Roma. [7]INFN Sezione di Napoli. [8]Max-Planck-Institut für Physik. [9]INFN Sezione di Padova and Università degli Studi di Padova. [10]Institut de Fisica d'Altes Energies (IFAE), The Barcelona Institute of Science and Technology. [11]LAPP, Univ. Grenoble Alpes, Univ. Savoie Mont Blanc, CNRS-IN2P3, Annecy. [12]Universität Hamburg, Institut für Experimentalphysik. [13]Graduate School of Science, University of Tokyo. [14]Universidad del Azuay. [15]INAF - Osservatorio di Astrofisica e Scienza dello spazio di Bologna. [16]Centro Brasileiro de Pesquisas Físicas. [17]Instituto de Astrofísica de Canarias and Departamento de Astrofísica, Universidad de La Laguna. [18]CIEMAT. [19]Department of Physics, TU Dortmund University. [20]INFN Sezione di Bari and Politecnico di Bari. [21]INAF - Osservatorio Astronomico di Brera. [22]INFN Sezione di Trieste and Università degli Studi di Udine. [23]University of Geneva - Départment de physique nucléaire et corpusculaire. [24]INFN Sezione di Catania. [25]INAF - Istituto di Astrofisica e Planetologia Spaziali (IAPS). [26]Aix Marseille Univ, CNRS/IN2P3, CPPM. [27]INFN Sezione di Bari and Università di Bari. [28]INFN Sezione di Torino. [29]Palacky University Olomouc, Faculty of Science. [30]University of Alcalá UAH. [31]Port d'Informació Científica. [32]University of Rijeka, Department of Physics. [33]Institute for Theoretical Physics and Astrophysics, Universität Würzburg. [34]Institut für Theoretische Physik, Lehrstuhl IV: Plasma-Astroteilchenphysik, Ruhr-Universität Bochum. [35]INFN Sezione di Roma La Sapienza. [36]ILANCE, CNRS . [37]Physics Program, Graduate School of Advanced Science and Engineering, Hiroshima University. [38]INFN Sezione di Roma Tor Vergata. [39]Faculty of Physics and Applied Informatics, University of Lodz. [40]University of Split, FESB. [41]Department of Physics, Yamagata University. [42]Josip Juraj Strossmayer University of Osijek, Department of Physics. [43]INFN Dipartimento di Scienze Fisiche e Chimiche - Università degli Studi dell'Aquila and Gran Sasso Science Institute. [44]Yukawa Institute for Theoretical Physics, Kyoto University. [45]FZU - Institute of Physics of the Czech Academy of Sciences. [46]Astronomical Institute of the Czech Academy of Sciences. [47]Faculty of Science, Ibaraki University. [48]Faculty of Science and Engineering, Waseda University. [49]Institute of Particle and Nuclear Studies, KEK (High Energy Accelerator Research Organization). [50]INFN Sezione di Trieste and Università degli Studi di Trieste. [51]INFN and Università degli Studi di Siena, Dipartimento di Scienze Fisiche, della Terra e dell'Ambiente (DSFTA). [52]Escuela Politécnica Superior de Jaén, Universidad de Jaén. [53]Saha Institute of Nuclear Physics. [54]Institute for Nuclear Research and Nuclear Energy, Bulgarian Academy of Sciences. [55]Dipartimento di Fisica e Chimica 'E. Segrè' Università degli Studi di Palermo. [56]Grupo de Electronica, Universidad Complutense de Madrid. [57]Hiroshima Astrophysical Science Center, Hiroshima University. [58]School of Allied Health Sciences, Kitasato University. [59]RIKEN, Institute of Physical and Chemical Re-

search. [60]Laboratory for High Energy Physics, École Polytechnique Fédérale. [61]Charles University, Institute of Particle and Nuclear Physics. [62]Division of Physics and Astronomy, Graduate School of Science, Kyoto University. [63]Institute for Space-Earth Environmental Research, Nagoya University. [64]Kobayashi-Maskawa Institute (KMI) for the Origin of Particles and the Universe, Nagoya University. [65]Graduate School of Technology, Industrial and Social Sciences, Tokushima University. [66]INFN Dipartimento di Scienze Fisiche e Chimiche - Università degli Studi dell'Aquila and Gran Sasso Science Institute. [67]Department of Physical Sciences, Aoyama Gakuin University. [68]IRFU, CEA, Université Paris-Saclay. [69]Department of Astronomy, University of Geneva. [70]Graduate School of Science and Engineering, Saitama University. [71]Institute of Space Sciences (ICE-CSIC), and Institut d'Estudis Espacials de Catalunya (IEEC), and Institució Catalana de Recerca I Estudis Avançats (ICREA). [72]Dipartimento di Fisica - Universitá degli Studi di Torino. [73]Department of Physics, Konan University.